\documentclass[12pt,twoside]{article}
\usepackage{amsfonts}
\usepackage{amsmath}
\usepackage{graphicx}
\usepackage[numbers,sort&compress]{natbib}

\date{}

\numberwithin{equation}{section}

\begin{document}

%\centerline{\bf Journal's Title, Vol. x, 20xx, no. xx, xxx - xxx}

%\centerline{}

%\centerline{}

\centerline{}

\centerline{\Large{\bf On the new wave behavior to the
longitudinal wave }} \centerline{\Large{\bf equation in a
magneto-electro-elastic circular rod }}

% \centerline{\Large{\bf Decomposition
%Method (NDM)}}

\centerline{}

\centerline{\bf {Onur Alp Ilhan$^1$, Hasan Bulut$^2$,}}

\centerline{\bf {Tukur A. Sulaiman$^3$ and Haci Mehmet
Baskonus$^4$}}

%\centerline{}

\centerline{$^{1}$Department of Mathematics, Erciyes University,
Kayseri, Turkey}

\centerline{$^{2,3}$Department of Mathematics, Firat University,
Elazig, Turkey}

\centerline{$^{3}$Department Mathematics, Federal University
Dutse, Jigawa, Nigeria}

\centerline{$^{4}$Department of Computer Engineering, Munzur
University, Tunceli, Turkey}

%\centerline{hbulut@firat.edu.tr$^1$, mtukur74@yahoo.com$^2$,
%hmbaskonus@gmail.com$^3$}

%\centerline{Address of Author1 third line}

%\centerline{Address of Author1 forth line}

%\centerline{}

%\centerline{\bf {}}

%\centerline{}

%\centerline{Firat University, Turkey}

%\centerline{Address of Author2 third line}

%\centerline{Address of Author2 forth line}
\centerline{}

%\centerline{\bf {}}

%\centerline{Firat University, Turkey} \centerline{}
\newtheorem{Theorem}{\quad Theorem}[section]

\newtheorem{Definition}[Theorem]{\quad Definition}

\newtheorem{Corollary}[Theorem]{\quad Corollary}

\newtheorem{Lemma}[Theorem]{\quad Lemma}

\newtheorem{Example}[Theorem]{\quad Example}

\centerline{}

\begin{abstract}
With the aid of the symbolic computations software; Wolfram
Mathematica 9, the powerful sine-Gordon expansion method is used
in examining the analytical solution of the longitudinal wave
equation in a magneto-electro-elastic circular rod. Sine-Gordon
expansion method is based on the well-known sine-Gordon equation
and a wave transformation. The longitudinal wave equation is an
equation that arises in mathematical physics with dispersion
caused by the transverse Poisson's effect in a
magneto-electro-elastic circular rod. We successfully get some
solutions with the complex, trigonometric and hyperbolic function
structure. We present the numerical simulations of all the
obtained solutions by choosing appropriate values of the
parameters. We give the physical meanings of some of the obtained
analytical solutions which significantly explain some practical
physical problems.
\end{abstract}

{\bf Keywords:} The SGEM; longitudinal wave equation in a MEE
circular rod; complex, hyperbolic, trigonometric function
solutions

%\twocolumn
\section{Introduction}
Searching for the new analytical solutions to nonlinear evolution
equations (NEEs) plays a vital role in the study of nonlinear
physical aspects. Nonlinear evolution equation are often used to
express complex models that arise in the various fields of
nonlinear sciences, such as; plasma physics, quantum mechanics,
biological sciences and so on. For the past two decades, various
analytical techniques have been invested to explore the search for
the new solutions to different type of NLEs such as the new
generalized algebra method \cite{blp:8}, the
tan($\frac{F(\xi)}{2}$)-expansion method \cite{pde:7, pde:8}, the
extended tanh method \cite{blp:7}, the jacobi elliptic function
method \cite{ref:6}, the homogeneous balance method \cite{mic:12},
the generalized Kudryashov method \cite{ref:5, ref:06}, the
generalized $(G^{'}/G)$ method \cite{ref:3}, the extended
homoclinic test function method \cite{pde:4}, the improved
Bernoulli sub-equation function method \cite{src:1}, the modified
exp $(-\Omega(\xi))$-expansion function method \cite{src:001,
src:2} and so on. Generally, many more analytical techniques have
been designed and used in obtaining analytical solutions of
various NLEs \cite{ref:7, ref:8, ref:9, ref:10, ref:11, blp:4,
blp:5, blp:6, ref:31, ref:22, ref:23, ref:24}.

In this work, the powerful sine-Gordon expansion method (SGEM)
\cite{sn:1, pde:12} is invested to search for some new solutions
to the longitudinal wave equation in a magneto-electro-elastic
(MEE) circular rod \cite{src:2}. The longitudinal wave equation is
an equation with dispersion caused by the transverse Poisson's
effect in a MEE circular rod which is derived by \cite{src:3}.

The longitudinal wave equation in a MEE circular rod is given by
\cite{src:2};

\begin{equation}\label{mee:1}
u_{tt}-v_{0}^{2}u_{xx}-\Big(\frac{v_{0}}{2}u^{2}+Mu_{tt}\Big)_{xx}=0,
\end{equation}
where $v_{0}$ is the linear longitudinal wave velocity for a MEE
circular rod and M is the dispersion parameter, all of them depend
on the material property and the geometry of the rod \cite{src:3}.
Various analytical approaches have been invested to seek for the
solutions of the longitudinal wave equation in a
magneto-electro-elastic MEE circular rod such as the modified
$(G^{'}/G)$-expansion method \cite{src:4},  the functional
variable method \cite{src:5}, the ansatz method \cite{src:6} and
so on.

\section{The SGEM}
In the present section, we give the general facts of SGEM.\\

Consider the following sine-Gordon equation \cite{{sn:2},{sn:3}}:

\begin{equation}\label{sng:1}
u_{xx}-u_{tt}=m^{2}sin(u).
\end{equation}
where $u=u(x,\;t)$ and $n\in \mathbb{R}\setminus \{0\}$.\\\\
Utilizing the wave transformation $u=u(x,\;t)=U(\zeta)$, $\zeta
=\alpha (x-kt)$ on Eq. (\ref{sng:1}), produces the following
nonlinear ordinary differential equation (NODE):
\begin{equation}\label{sng:2}
U^{''}=\frac{n^{2}}{\alpha^{2}(1-k^{2})}sin(U),
\end{equation}
where $U=U(\zeta)$, $\zeta$ is the amplitude of the travelling
wave and $k$ is the speed of the travelling wave. Integrating Eq.
(\ref{sng:2}), we obtain the following equation:

\begin{equation}\label{sng:3}
\bigg[\Big(\frac{U}{2}\Big)^{'}
\bigg]^{2}=\frac{n^{2}}{\alpha^{2}(1-k^{2})}sin^{2}\Big(\frac{U}{2}\Big)+Q,
\end{equation}
where $Q$ is the integration constant.\\

Substituting $Q=0$, $\omega(\zeta)=\frac{U}{2}$ and
$b^{2}=\frac{n^{2}}{\alpha^{2}(1-k^{2})}$ in Eq. (\ref{sng:3}),
gives:

\begin{equation}
\label{sng:4} \omega^{'}=b sin(\omega),
\end{equation}
inserting $b=1$ into Eq. (\ref{sng:4}), produces:

\begin{equation}
\label{sng:5} \omega^{'}= sin(\omega),
\end{equation}
simplifying Eq. (\ref{sng:5}), gives the following two significant
equations;

\begin{equation}
\label{sng:6}
sin(\omega)=sin(\omega(\zeta))=\frac{2de^{\zeta}}{d^{2}e^{2\zeta}+1}\Bigg
|_{d=1}=sech(\zeta),
\end{equation}

\begin{equation}
\label{sng:7}
cos(\omega)=cos(\omega(\zeta))=\frac{d^{2}e^{2\zeta}-1}{d^{2}e^{2\zeta}+1}\Bigg
|_{d=1}=tanh(\zeta),
\end{equation}
where $d$ is the integral constant.\\

For the given nonlinear partial differential equation Eq.
(\ref{sng:8});

\begin{equation}
\label{sng:8} P(u,\;uu_{x},\;u^{2}u_{t},\; \ldots),
\end{equation}
we look its solution in the form;

\begin{equation}
\label{sng:9}
U(\zeta)=\sum_{i=1}^{m}tanh^{i-1}(\zeta)\big[B_{i}sech(\zeta)+A_{i}tanh(\zeta)\big]+A_{0}.
\end{equation}
Equation (\ref{sng:9}) may be given according to Eq. (\ref{sng:6})
and (\ref{sng:7}) as;

\begin{equation}
\label{sng:10}
U(\omega)=\sum_{i=1}^{m}cos^{i-1}(\omega)\big[B_{i}sin(\omega)+A_{i}cos(\omega)\big]+A_{0}.
\end{equation}
We determine $m$ by balancing the highest power nonlinear term and
the highest derivative in the transformed NODE. Taking each
summation of the coefficients of $sin^{i}(w)cos^{j}(w)$, $0\le i,
j\le m$ to be zero, produces a set of equations. Solving this set
of equation with the symbolic computational software like Wolfram
Mathematica 9, yields the values of the coefficients
$A_{i},\;B_{i},\;\mu$ and $c$. Finally, inserting the obtained
values of these coefficients into Eq. (\ref{sng:9}) along with the
value of $m$, gives the new travelling wave solutions to Eq.
(\ref{sng:8}).

\section{Applications}
In this section, the SGEM is used in searching the new solutions
to Eq. (\ref{mee:1}). Considering Eq. (\ref{mee:1}), we derive the
following NODE by utilizing the wave transformation; $u=U(\zeta)$,
$\zeta=\mu(x-kt)$;

\begin{equation}\label{mee:2}
2pk^{2}\mu^{2}U^{''}-2(k^{2}-c_{0}^{2})U+c_{0}^{2}U^{2}=0,
\end{equation}
we get $m=2$ by balancing $U^{''}$ and $U^{2}$ in Eq.
(\ref{mee:2}).\\\\
Using Eq. (\ref{sng:10}) together with the value $m=2$, we get the
following equation;

\begin{equation}
\label{mee:3}
U(\omega)=B_{1}sin(\omega)+A_{1}cos(\omega)+B_{2}cos(\omega)sin(\omega)+A_{2}cos^{2}(\omega)+A_{0},
\end{equation}
differentiating Eq. (\ref{mee:3}) twice, we get:

\begin{equation}\label{mee:4}
\begin{split}
U^{''}(\omega)=B_{1}cos^{2}(\omega)sin(\omega)-B_{1}sin^{3}(\omega)-2A_{1}sin^{2}(\omega)cos(\omega)\\+B_{2}cos^{3}(\omega)sin(\omega)
-5B_{2}sin^{3}(\omega)cos(\omega)-4A_{2}cos^{2}(\omega)sin^{\omega}(\omega)+2A_{2}sin^{4}(\omega),
\end{split}
\end{equation}
Putting Eq. (\ref{mee:3}) and (\ref{mee:4}) into Eq.
(\ref{mee:2}), yields an equation in trigonometric functions.
After making some trigonometric identities substitutions into the
trigonometric functions equation, we collect a set of algebraic
equations by setting each summation of the coefficients of the
trigonometric functions of the same power to zero. We solve the
set of equations with the aid of the symbolic software;
Mathematica or Maple to get the various cases for the values of
the coefficients. We insert the values of the coefficients for
each case into Eq. (\ref{sng:9}) along with the value $m=2$, this
gives us new solution to Eq. Eq. (\ref{mee:1}).\\\\

\textbf{Case-1:}

$A_{0}=\frac{4}{c_{0}^{2}}(c_{0}^{2}-k^{2}), A_{1}=0, B_{1}=0,
A_{2}=-\frac{6}{c_{0}^{2}}(c_{0}^{2}-k^{2}),
B_{2}=6\Big(\frac{k^{2}}{c_{0}^{2}}-1\Big)i,
\\p=\frac{1}{k^{2}\mu^{2}}(k^{2}-c_{0}^{2}).$\\

\textbf{Case-2:}

$A_{0}=4\Big(1-\frac{1}{1+p\mu^{2}}\Big), A_{1}=0, B_{1}=0,
A_{2}=6\Big(\frac{1}{1+p\mu^{2}}-1\Big),
B_{2}=\frac{6p\mu^{2}(p\mu^{2}-1)}{p^{2}\mu^{4}-1}i,
\\c_{0}=-k\sqrt{1+p\mu^{2}}.$\\

\textbf{Case-3:}

$A_{0}=6\Big(\frac{k^{2}}{c_{0}^{2}}-1\Big), A_{1}=0, B_{1}=0,
A_{2}=6\Big(1-\frac{k^{2}}{c_{0}^{2}}\Big),
B_{2}=6\Big(1-\frac{k^{2}}{c_{0}^{2}}\Big)i,
\\\mu=-\frac{1}{k\sqrt{p}}(k^{2}-c_{0}^{2}).$\\

\textbf{Case-4:}
$$A_{0}=\frac{1}{c_{0}^{2}}(c_{0}^{2}-k^{2}), A_{1}=0, B_{1}=0,
A_{2}=-\frac{3}{c_{0}^{2}}(c_{0}^{2}-k^{2}), B_{2}=0,
p=\frac{k^{2}-c_{0}^{2}}{4k^{2}\mu^{2}}.$$

\textbf{Case-5:}

$A_{0}=1-\frac{1}{4p\mu^{2}+1}, A_{1}=0, B_{1}=0,
A_{2}=3\Big(\frac{1}{4p\mu^{2}+1}-1\Big), B_{2}=0,
\\c_{0}=k\sqrt{4p\mu^{2}+1}.$

\textbf{Case-6:}

$$A_{0}=1-\frac{k^{2}}{c_{0}^{2}}, A_{1}=0, B_{1}=0,
A_{2}=3\Big(\frac{k^{2}}{c_{0}^{2}}-1\Big), B_{2}=0,
\mu=\frac{1}{2k\sqrt{p}}(k^{2}-c_{0}^{2})i.$$

With case-1, we get the following solution;

\begin{equation}\label{mee:4}
\begin{split}
u_{1}(x,
t)=\frac{(k^{2}-c_{0}^{2})}{c_{0}^{2}}\Big(6+6i\;sech[\mu(x-kt)]tanh[\mu(x-kt)]\\-6tanh^{2}[\mu(x-kt)]\Big).
\end{split}
\end{equation}

\begin{figure}[h]
\begin{center}$
\begin{array}{cc}
\includegraphics[width=50mm]{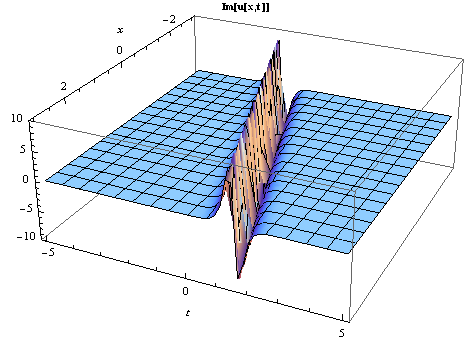}&
\includegraphics[width=50mm]{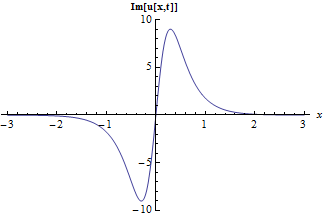}
\end{array}$
\end{center}
\caption{The 3D and 2D shape for the imaginary part of Eq.
(\ref{mee:4}) with the values $k=2$, $c_{0}=1$, $\mu=3$, $-3<x<3$,
$-5<t<5$ and $t=0$ for the 2D graphic.} \label{pics:1}
\end{figure}
\newpage
\begin{figure}[h]
\begin{center}$
\begin{array}{cc}
\includegraphics[width=50mm]{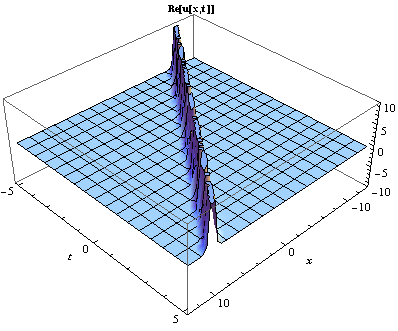}&
\includegraphics[width=50mm]{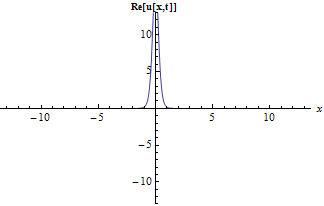}
\end{array}$
\end{center}
\caption{The 3D and 2D shape for the real part of Eq.
(\ref{mee:4}) with the values $k=2$, $c_{0}=1$, $\mu=3$,
$-13<x<13$, $-5<t<5$ and $t=0$ for the 2D graphic.} \label{pics:2}
\end{figure}

With case-2, we get the following solution;

\begin{equation}\label{mee:5}
\begin{split}
u_{2}(x,
t)=\Big(1-\frac{1}{1+p\mu^{2}}\Big)\Big(4-6tanh^{2}[\mu(x-kt)]\Big)\\+
\frac{1}{p^{2}\mu^{2}-1}\Big(6p\mu^{2}(p\mu^{2}-1)i\;sech[\mu(x-kt)]tanh[\mu(x-kt)]\Big).
\end{split}
\end{equation}
\newpage
\begin{figure}[h]
\begin{center}$
\begin{array}{cccc}
\includegraphics[width=50mm]{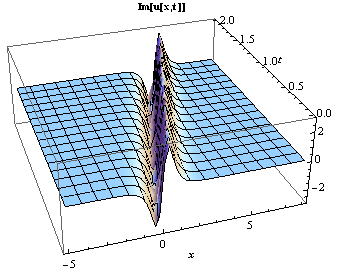}&
\includegraphics[width=50mm]{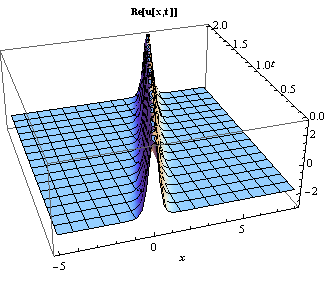}&\\
\includegraphics[width=50mm]{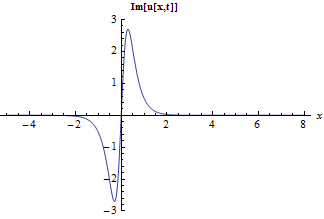}&
\includegraphics[width=50mm]{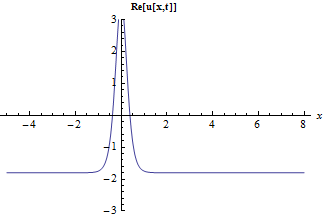}
\end{array}$
\end{center}
\caption{The 3D and 2D shape for the real part of Eq.
(\ref{mee:5}) with the values $k=2$, $p=1$, $\mu=3$, $-5<x<8$,
$0<t<2$ and $t=0$ for the 2D graphics.} \label{pics:3}
\end{figure}
%\newpage
With case-3, we get the following solution;

\begin{equation}\label{mee:6}
\begin{split}
u_{3}(x,
t)=\frac{1}{c_{0}^{2}}(c_{0}^{2}-k^{2})\Big(-1-i\;sech\Big[\frac{1}{k\sqrt{p}}(k^{2}-c_{0}^{2})(x-kt)\Big]\\
\times
tanh\Big[\frac{1}{k\sqrt{p}}(k^{2}-c_{0}^{2})(x-kt)\Big]+tanh^{2}\Big[\frac{1}{k\sqrt{p}}(k^{2}-c_{0}^{2})(x-kt)\Big]\Big).
\end{split}
\end{equation}
\newpage
\begin{figure}[h]
\begin{center}$
\begin{array}{cccc}
\includegraphics[width=50mm]{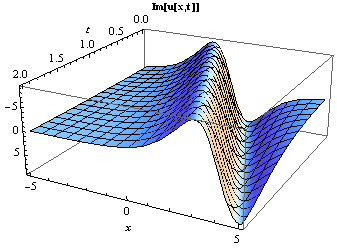}&
\includegraphics[width=50mm]{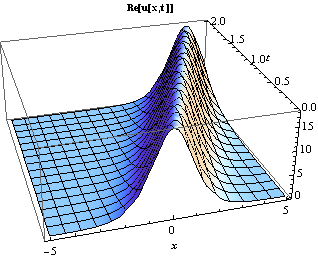}&\\
\includegraphics[width=50mm]{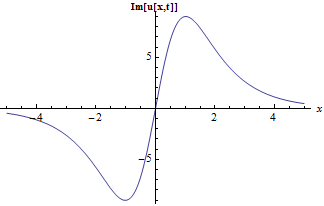}&
\includegraphics[width=50mm]{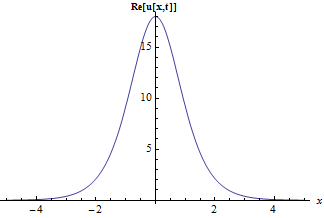}
\end{array}$
\end{center}
\caption{The 3D and 2D shape for the real part of Eq.
(\ref{mee:6}) with the values $k=2$, $p=1$, $c_{0}=1$, $-5<x<5$,
$0<t<2$ and $t=0$ for the 2D graphics.} \label{pics:4}
\end{figure}

With case-4, we get the following solution;

\begin{equation}\label{mee:7}
\begin{split}
u_{4}(x,
t)=\frac{1}{c_{0}^{2}}\Big((c_{0}^{2}-k^{2})-3(c_{0}^{2}-k^{2})tanh^{2}[\mu(x-kt)]\Big).
\end{split}
\end{equation}

\begin{figure}[h]
\begin{center}$
\begin{array}{cccc}
\includegraphics[width=50mm]{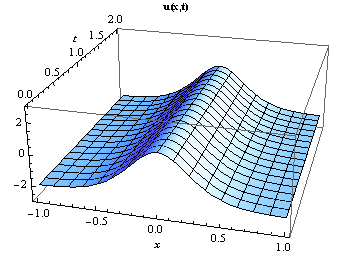}&
\includegraphics[width=50mm]{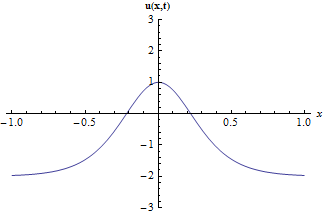}
\end{array}$
\end{center}
\caption{The 3D and 2D shape for the real part of Eq.
(\ref{mee:7}) with the values $k=0.005$, $\mu=3$, $c_{0}=1$,
$-1<x<1$, $0<t<2$ and $t=0$ for the 2D graphic.} \label{pics:5}
\end{figure}

With case-5, we get the following solution;

\begin{equation}\label{mee:8}
\begin{split}
u_{5}(x,
t)=\frac{4p\mu^{2}}{1+4p\mu^{2}}\Big(1-3tanh^{2}[\mu(x-kt)]\Big).
\end{split}
\end{equation}

\begin{figure}[h]
\begin{center}$
\begin{array}{cccc}
\includegraphics[width=50mm]{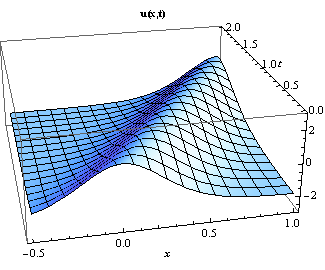}&
\includegraphics[width=50mm]{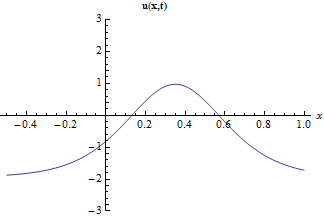}
\end{array}$
\end{center}
\caption{The 3D and 2D shape for the real part of Eq.
(\ref{mee:8}) with the values $k=0.5$, $\mu=3$, $p=1$, $-0.5<x<1$,
$0<t<2$ and $t=0.7$ for the 2D graphic.} \label{pics:6}
\end{figure}

With case-6, we get the following solution;

\begin{equation}\label{mee:9}
\begin{split}
u_{6}(x,
t)=-\frac{k^{2}-c_{0}^{2}}{c_{0}^{2}}\Big(1+3tan^{2}\Big[\frac{\sqrt{k^{2}-c_{0}^{2}}}{2k\sqrt{p}}(x-kt)\Big]\Big).
\end{split}
\end{equation}

\begin{figure}[h]
\begin{center}$
\begin{array}{cccc}
\includegraphics[width=50mm]{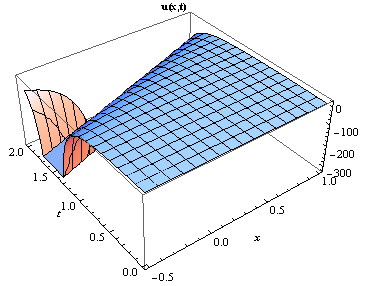}&
\includegraphics[width=50mm]{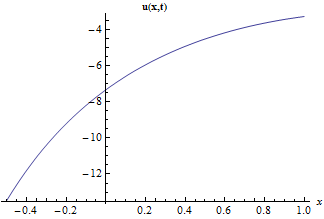}
\end{array}$
\end{center}
\caption{The 3D and 2D shape for the real part of Eq.
(\ref{mee:9}) with the values $k=2$, $c_{0}=1$, $p=1$, $-0.5<x<1$,
$0<t<2$ and $t=0.7$ for the 2D graphic.} \label{pics:7}
\end{figure}
\newpage
\section{Discussion and Remarks}
In \cite{src:2} the modified exp $(-\Omega(\xi))$-expansion
function method was developed and been utilized in solving the
longitudinal wave equation in a magneto-electro-elastic circular
rod and various solutions in hyperbolic functions form were
obtained. Secondly, the well-known modified $(G^{'}/G)$-expansion
method \cite{src:4} has been employed to this equation and some
exact hyperbolic and trigonometric function were obtained. We
observe that our results are new, but with the same solution
structures when compared with the existing results obtained by
using these two methods. On the other hand, we observe that in the
numerical simulations of the solutions we presented; fig. 1, 2, 7
are singular soliton surfaces, fig. 3 is solitoff surface, fig. 4,
5, 6 are soliton surfaces. We observed that some solutions in this
study have important physical meanings, like the hyperbolic
tangent arises in the calculation of magnetic moment and rapidity
of special relativity and the hyperbolic secant arises in the
profile of a laminar jet \cite{cln:2}.

\section{Conclusions}
In this study, by utilizing the sine-Gordon expansion method with
the help of Wolfram Mathematica 9, we investigated the solutions
of the longitudinal wave equation in a magneto-electro-elastic
circular rod. We obtained some new complex hyperbolic and
trigonometric function solutions. All the obtained solutions in
this study verified the longitudinal wave equation in a
magneto-electro-elastic circular rod, we checked this using the
same program in Wolfram Mathematica 9. We performed the numerical
simulations of all the obtained solutions in this article. We
observed that our results might be helpful in detecting the
transverse Poisson's effect in a magneto-electro-elastic circular
rod. Sine-Gordon expansion method is powerful and efficient
mathematical tool that can be used with the aid of symbolic
software such as Maple or Mathematica in exploring search for the
solutions of the various nonlinear equations arising in the
various field of nonlinear sciences.

%\newpage


\begin{thebibliography}{99}
%%%%%%%%%%%%%%%%%%%%%%%%%%%%%%%%%%%%%%%%%%%%%%%%%%%%%%%%%%%%%%%%%%%%%%%%%%%%%%%%%%%%%%%%%%%%%%%%%%%%%%%%%%%%%%%%%%%%%%%%%%%%%%%%%%%%%%%
\bibitem{blp:8} Y.J. Ren, S.T. Liu and H.Q. Zhang, \emph{A New Generalized Algebra Method and its Application in
the (2+1)-Dimensional Boiti-Leon-Pempinelli Equation}, Chaos,
Solitons and Fractals, 32 (2007) 1655-1665
%%%%%%%%%%%%%%%%%%%%%%%%%%%%%%%%%%%%%%%%%%%%%%%%%%%%%%%%%%%%%%%%%%%%%%%%%%%%%%%%%%%%%%%%%%%%%%%%%%%%%%%%%%%%%%%%%%%%%%%%%%%%%%%%%%%%
\bibitem{pde:7} J. Manafian, \emph{Optical Soliton Solutions for Schrodinger Type Nonlinear Evolution Equations by the tan($\frac{F(\xi)}{2}$)-Expansion Method},
Optik-International Journal of Light and Electron Optics, 127
(2016) 4222-4245
%%%%%%%%%%%%%%%%%%%%%%%%%%%%%%%%%%%%%%%%%%%%%%%%%%%%%%%%%%%%%%%%%%%%%%%%%%%%%%%%%%%%%%%%%%%%%%%%%%%%%%%%%%%%%%%%%%%%%%%%%%%%%%%%%%%%%%%
\bibitem{pde:8} Y. Ugurlu, I.E. Inan and H. Bulut, \emph{Two New Application of tan($\frac{F(\xi)}{2}$)-Expansion Method},
Optik-International Journal of Light and Electron Optics, 131
(2017) 539-536
%%%%%%%%%%%%%%%%%%%%%%%%%%%%%%%%%%%%%%%%%%%%%%%%%%%%%%%%%%%%%%%%%%%%%%%%%%%%%%%%%%%%%%%%%%%%%%%%%%%%%%%%%%%%%%%%%%%%%%%%%%%%%%%%%%%%%%%
\bibitem{blp:7} W.G. Feng, K.M. Li, Y.Z. Li and C. Lin, \emph{Explicit Exact Solutions for (2+1)-Dimensional Boiti-Leon-Pempinelli
Equation}, Commun Nonlinear Sci Numer Simulat, 14 (2009) 2013-2017
%%%%%%%%%%%%%%%%%%%%%%%%%%%%%%%%%%%%%%%%%%%%%%%%%%%%%%%%%%%%%%%%%%%%%%%%%%%%%%%%%%%%%%%%%%%%%%%%%%%
\bibitem{ref:6} Z. Fu, S. Liu, Q. Zhao, \emph{New Jacobi elliptic function expansion and new periodic solutions of nonlinear wave equations},
Physics Letters A 290 (2001) 72-76
%%%%%%%%%%%%%%%%%%%%%%%%%%%%%%%%%%%%%%%%%%%%%%%%%%%%%%%%%%%%%%%%%%%%%%%%%%%%%%%%%%%%%%%%%%%%%%%%%%%%%%%%%%%%%%%%%%%%%%%%%%%%%%%%%%%%%%%
\bibitem{mic:12} E.M.E. Zayed and K.A.E. Alurrfi, \emph{The Homogeneous Balance Method and its Applications for
Finding the Exact Solutions for Nonlinear Evolution Equations},
Italian Journal of Pure and Applied Mathematics, 33 (2014) 307-318
%%%%%%%%%%%%%%%%%%%%%%%%%%%%%%%%%%%%%%%%%%%%%%%%%%%%%%%%%%%%%%%%%%%%%%%%%%%%%%%%%%%%%%%%%%%%%%%%%%%%%%%%%%%%%%%%%%%%%%%%%%%%%%%%%%%%%%%
\bibitem{ref:5} M.S. Islam, K. Khan and A.H. Arnous, \emph{Generalized Kudryashov Method for Solving Some
(3+1)-Dimensional Nonlinear Evolution Equations}, New Trends in
Mathematical Sciences, 3(3) (2015) 46-57
%%%%%%%%%%%%%%%%%%%%%%%%%%%%%%%%%%%%%%%%%%%%%%%%%%%%%%%%%%%%%%%%%%%%%%%%%%%%%%%%%%%%%%%%%%%%%%%%%%%%%%%%%%%
\bibitem{ref:06} M. Kaplan, A. Bekir and A. Akbulut, \emph{A Generalized Kudryashov Method to Some Nonlinear Evolution Equations
in Mathematical Physics}, Nonlinear Dynamics, 85(4) (2016)
2843-2850
%%%%%%%%%%%%%%%%%%%%%%%%%%%%%%%%%%%%%%%%%%%%%%%%%%%%%%%%%%%%%%%%%%%%%%%%%%%%%%%%%%%%%%%%%%%%%%%%%%%%%%%%%%%
\bibitem{ref:3} H.L. Lu, X.Q. Liu, L. Niu, A generalized $(G^{'}/G)$-expansion method and its applications to nonlinear evolution equations,
 Applied Mathematics and Computation 215 (2010) 3811-3816
%%%%%%%%%%%%%%%%%%%%%%%%%%%%%%%%%%%%%%%%%%%%%%%%%%%%%%%%%%%%%%%%%%%%%%%%%%%%%%%%%%%%%%%%%%%%%%%%%%%%%%%%%%%%%%%%%%%%%%%%%%%%%%%%%%%%%%%
\bibitem{pde:4} C. Wang, \emph{Dynamic Behavior of Traveling Waves for the Sharma-Tasso-Olver Equation}, Nonlinear Dynamics, 85(2) (2016), 1119-1126.
%%%%%%%%%%%%%%%%%%%%%%%%%%%%%%%%%%%%%%%%%%%%%%%%%%%%%%%%%%%%%%%%%%%%%%%%%%%%%%%%%%%%%%%%%%%%%%%%%%%%%%%%%%%%%%%%%%%%%%%
\bibitem{src:1} H.M. Baskonus and H. Bulut \emph{Exponential prototype structures for (2+1)-dimensional
Boiti-Leon-Pempinelli systems in mathematical physics}, Waves in
Random and Complex Media, 26(2) (2016) 201-208
%%%%%%%%%%%%%%%%%%%%%%%%%%%%%%%%%%%%%%%%%%%%%%%%%%%%%%%%%%%%%%%%%%%%%%%%%%%%%%%%%%%%%%%%%%%%%%%%%%%%%%%%%%%%%%%%%%%%%%%%%%%%%%%%%%%%%%%
\bibitem{src:001} H.M. Baskonus, H. Bulut and F.B.M. Belgacem, \emph{Analytical Solutions for Nonlinear Long-Short Wave Interaction
Systems with Highly Complex Structure}, Journal of Computational
and Applied Mathematics, 312 (2017) 257-266
%%%%%%%%%%%%%%%%%%%%%%%%%%%%%%%%%%%%%%%%%%%%%%%%%%%%%%%%%%%%%%%%%%%%%%%%%%%%%%%%%%%%%%%%%%%%%%%%%%%%%%%%%%%%%%%%%%%%%%%%%%%%%%%%%%%%%%%
\bibitem{ref:7} N. Kadkhoda, H. Jafari, \emph{Kudryashov Method for Exact
Solutions of Isothermal Magnetostatic Atmospheres},
 Applied Mathematics and Computation, 6(1) (2016) 43-52
%%%%%%%%%%%%%%%%%%%%%%%%%%%%%%%%%%%%%%%%%%%%%%%%%%%%%%%%%%%%%%%%%%%%%%%%%%%%%%%%%%%%%%%%%%%%%%%%%%%%%%%%%%%%%%%%%%%%%%%%%%%%%%%%%%%%%%%
\bibitem{ref:8} S.A. Elwakil, S.K. El-labany, M.A. Zahran, R. Sabry,
\emph{Modified extended tanh-function method for solving nonlinear
partial differential equations}, Physics Letters A  299 (2002)
179-188
%%%%%%%%%%%%%%%%%%%%%%%%%%%%%%%%%%%%%%%%%%%%%%%%%%%%%%%%%%%%%%%%%%%%%%%%%%%%%%%%%%%%%%%%%%%%%%%%%%%%%%%%%%%%%%%%%%%%%%%%%%%%%%%%%%%%%%%
\bibitem{ref:9} M.S. Islam, K. Khan, A.H. Arnous, \emph{Generalized Kudryashov Method for Solving Some
(3+1)-Dimensional Nonlinear Evolution Equations}, New Trends in
Mathematical Sciences, 3(3) (2015) 46-57
%%%%%%%%%%%%%%%%%%%%%%%%%%%%%%%%%%%%%%%%%%%%%%%%%%%%%%%%%%%%%%%%%%%%%%%%%%%%%%%%%%%%%%%%%%%%%%%%%%%%%%%%%%%%%%%%%%%%%%%%%%%%%%%%%%%%%%%
\bibitem{ref:10} H.T. Chen, Z. Hong-Qing, \emph{New double periodic and multiple soliton solutions of the generalized (2 + 1)-dimensional Boussinesq equation},
Chaos, Solitons and Fractals 20 (2004) 765-769
%%%%%%%%%%%%%%%%%%%%%%%%%%%%%%%%%%%%%%%%%%%%%%%%%%%%%%%%%%%%%%%%%%%%%%%%%%%%%%%%%%%%%%%%%%%%%%%%%%%%%%%%%%%%%%%%%%%%%%%%%%%%%%%%%%%%%%%
\bibitem{ref:11} H.M Baskonus, H. Bulut, \emph{On Some New Analytical Solutions for The (2+1)-Dimensional Burgers Equation And The Special
Type Of Dodd-Bullough-Mikhailov Equation}, Journal of Applied
Analysis and Computation 5(4) (2015) 613-625
%%%%%%%%%%%%%%%%%%%%%%%%%%%%%%%%%%%%%%%%%%%%%%%%%%%%%%%%%%%%%%%%%%%%%%%%%%%%%%%%%%%%%%%%%%%%%%%%%%%%%%%%%%%%%%%%%%%%%%%%%%%%%%%%%%%%%%%
\bibitem{blp:4} Z. Lu and H. Zhang, \emph{Soliton Like and Multi-Soliton Like Solutions for the
Boiti-Leon-Pempinelli Equation}, Chaos, Solitons and Fractals, 19
(2004) 527-531
%%%%%%%%%%%%%%%%%%%%%%%%%%%%%%%%%%%%%%%%%%%%%%%%%%%%%%%%%%%%%%%%%%%%%%%%%%%%%%%%%%%%%%%%%%%%%%%%%%%
\bibitem{blp:5} D.J. Huang and H.Q. Zhang, \emph{Exact Travelling Wave Solutions for
the Boiti-Leon-Pempinelli Equation}, Chaos, Solitons and Fractals,
22 (2004) 243-247
%%%%%%%%%%%%%%%%%%%%%%%%%%%%%%%%%%%%%%%%%%%%%%%%%%%%%%%%%%%%%%%%%%%%%%%%%%%%%%%%%%%%%%%%%%%%%%%%%%%
\bibitem{blp:6} C. Dai and Y. Wang, \emph{Periodic Structures Based on Variable Separation Solution of the
(2+1)-Dimensional Boiti-Leon-Pempinelli Equation}, Chaos, Solitons
and Fractals, 39 (2009) 350-355

\bibitem{ref:31} Y. Liang, \emph{Exact Solutions of the (3+1)-Dimensional Modified KdV-Zakharov-Kuznetsev equation and Fisher equations using the modified Simple
Equation Method}, Journal of Interdisciplinary Mathematics, 17
(2014) 565-578
%%%%%%%%%%%%%%%%%%%%%%%%%%%%%%%%%%%%%%%%%%%%%%%%%%%%%%%%%%%%%%%%%%%%%%%%%%%%%%%%%%%%%%%%%%%%%%%%%%%%%%%%%%%%%%%%%%%%%%%
\bibitem{ref:22} H. Zhang, \emph{Extended Jacobi Elliptic Function Expansion Method and its Applications},
Communications in Nonlinear Science and Numerical Simulation,
12(5) (2007)  627-635
%%%%%%%%%%%%%%%%%%%%%%%%%%%%%%%%%%%%%%%%%%%%%%%%%%%%%%%%%%%%%%%%%%%%%%%%%%%%%%%%%%%%%%%%%%%%%%%%%%%%%%%%%%%%%%%%%%%%%%%%%%%%%%%%%%%%%%
\bibitem{ref:23} N.Z. Petrovic, M. Bohra, \emph{General Jacobi Elliptic Function Expansion Method Applied to the Generalized
(3 + 1)-Dimensional Nonlinear Schrodinger Equation}, Optical and
Quantum Electronics, 48(268) (2016)
%%%%%%%%%%%%%%%%%%%%%%%%%%%%%%%%%%%%%%%%%%%%%%%%%%%%%%%%%%%%%%%%%%%%%%%%%%%%%%%%%%%%%%%%%%%%%%%%%%%%%%%%%%%%%%%%%%%%%%%%%%%%%%%%%%%%%%
\bibitem{ref:24} Z. Yan, \emph{Jacobi Elliptic Function Solutions of Nonlinear Wave
Equations via the New sinh-Gordon Equation Expansion Method}, MM
Research Preprints, 22 (2003) 363-375
%%%%%%%%%%%%%%%%%%%%%%%%%%%%%%%%%%%%%%%%%%%%%%%%%%%%%%%%%%%%%%%%%%%%%%%%%%%%%%%%%%%%%%%%%%%%%%%%%%%%%%%%%%%%%%%%%%%%%%%%%%%%%%%%%%%%%%
%%%%%%%%%%%%%%%%%%%%%%%%%%%%%%%%%%%%%%%%%%%%%%%%%%%%%%%%%%%%%%%%%%%%%%%%%%%%%%%%%%%%%%%%%%%%%%%%%%%%%%%%%%%%%%%%%%%%%%%%%%%%%%%%%%%%%%%%%%%%%%%%%%%%%%
\bibitem{sn:1} C. Yan, \emph{A Simple Transformation for Nonlinear Waves}, Physics Letters A, 22(4) (1996) 77-84
%%%%%%%%%%%%%%%%%%%%%%%%%%%%%%%%%%%%%%%%%%%%%%%%%%%%%%%%%%%%%%%%%%%%%%%%%%%%%%%%%%%%%%%%%%%%%%%%%%%%%%%%%%%%%%%%%%%%%%%%%%%%%%%%%%%%%
\bibitem{pde:12} H. Bulut, T.A. Sulaiman and H.M. Baskonus, \emph{New Solitary and Optical Wave Structures
to the Korteweg-de Vries Equation with Dual-Power Law
Nonlinearity}, Opt Quant Electron, 48(564) (2016) 1-14
%%%%%%%%%%%%%%%%%%%%%%%%%%%%%%%%%%%%%%%%%%%%%%%%%%%%%%%%%%%%%%%%%%%%%%%%%%%%%%%%%%%%%%%%%%%%%%%%%%%%%%%%%%%%%%%%%%%%%%%
\bibitem{src:2}  H.M. Baskonus, H. Bulut and Abdon Atangana, \emph{On the Complex and Hyperbolic Structures of Longitudinal
 Wave Equation in a Magneto-Electro-Elastic Circular Rod}, Smart Materials and Structures, 25(3) (2016), 035022
 %%%%%%%%%%%%%%%%%%%%%%%%%%%%%%%%%%%%%%%%%%%%%%%%%%%%%%%%%%%%%%%%%%%%%%%%%%%%%%%%%%%%%%%%%%%%%%%%%%%%%%%%%%%%%%%%%%%
 \bibitem{src:3}  C.X. Xue, E. Pan and X.Y. Zhang, \emph{Solitary Waves in a Magneto-Electro-Elastic
Circular Rod}, Smart Materials and Structures, 20(10) (2011),
035022
 %%%%%%%%%%%%%%%%%%%%%%%%%%%%%%%%%%%%%%%%%%%%%%%%%%%%%%%%%%%%%%%%%%%%%%%%%%%%%%%%%%%%%%%%%%%%%%%%%%%%%%%%%%%%%%%%%%%
\bibitem{src:4}  X. Ma, Y. Pan and L. Chang, \emph{Explicit Travelling Wave Solutions in a Magneto-Electro-Elastic
Circular Rod}, International Journal of Computer Science Issues,
10(1) (2013), 62-68
 %%%%%%%%%%%%%%%%%%%%%%%%%%%%%%%%%%%%%%%%%%%%%%%%%%%%%%%%%%%%%%%%%%%%%%%%%%%%%%%%%%%%%%%%%%%%%%%%%%%%%%%%%%%%%%%%%%%
\bibitem{src:5}  K. Khan, H. Koppelaar and A. Akbar, \emph{Exact and Numerical Soliton Solutions to Nonlinear Wave Equations},
Computational and Mathematical Engineering, 2 (2016), 5-22
 %%%%%%%%%%%%%%%%%%%%%%%%%%%%%%%%%%%%%%%%%%%%%%%%%%%%%%%%%%%%%%%%%%%%%%%%%%%%%%%%%%%%%%%%%%%%%%%%%%%%%%%%%%%%%%%%%%%
\bibitem{src:6}  M. Younis and S. Ali, \emph{Bright, Dark, and Singular Solitons in Magneto-Electro-Elastic Circular Rod},
Waves in Random and Complex Media, 25(4) (2015), 549-555
 %%%%%%%%%%%%%%%%%%%%%%%%%%%%%%%%%%%%%%%%%%%%%%%%%%%%%%%%%%%%%%%%%%%%%%%%%%%%%%%%%%%%%%%%%%%%%%%%%%%%%%%%%%%%%%%%%%%
 \bibitem{sn:2} Z. Yan and  H. Zhang, \emph{New Explicit and exact Travelling Wave Solutions for a System of Variant Boussinesq equations
in Mathematical Physics}, Physics Letters A, 252 (1999), 291-296

%%%%%%%%%%%%%%%%%%%%%%%%%%%%%%%%%%%%%%%%%%%%%%%%%%%%%%%%%%%%%%%%%%%%%%%%%%%%%%%%%%%%%%%%%%%%%%%%%%%%%%%%%%%%%%%%%%%%%%%%%%%%%%%%%%%%%%%
\bibitem{sn:3} Y. Zhen-Ya, Z. Hong-oing and F. En-Gui, \emph{New Explicit and Travelling Wave Solutions
for a Class of Nonlinear Evolution Equations}, Acta Physica
Sinica, 48(1) (1999), 1-5

 \bibitem{cln:2} E.W. Weisstein,  \emph{Concise Encyclopedia of Mathematics}, 2$^nd$ edition.
New York: CRC Press,
      (2002).

\end{thebibliography}
\end{document}